\documentclass[12pt]{article}
\usepackage[english]{babel}
\usepackage{epsfig,graphicx}
\usepackage{amsmath,amssymb}
 
\usepackage[nomarkers,nolists]{endfloat}
\makeatletter
\let\table\@btab
\makeatother
 
\def\etal{et al.\@}
\def\vs{--}
 
\def\PM#1#2{\,{}^{+\,#1}_{-\,#2}}
\def\deg{^\circ}
 
\def\ds{{D^*}}
\def\dspm{{D^{*\pm}}}
\def\dss{{D_s}}
\def\dsspm{{D_s^\pm}}
 
\def\WJB{W_{\textrm{JB}}}

\def\pt{p_{\!\perp}}
 
\def\ptdss{\pt^{\dss}}
\def\gs{\gamma_s}
 
\def\perc{\,\%}
\def\GeV{\,\textrm{GeV}}
\def\MeV{\,\textrm{MeV}}
\def\g2{\GeV^2}
\def\nb{\,\textrm{nb}}
\def\pb{\,\textrm{pb}}
\def\ipb{\pb^{-1}}
\def\stat{\,\textrm{(stat.)}}
\def\syst{\,\textrm{(syst.)}}
\def\br{\,\textrm{(br.)}}
\def\frcm{\,\textrm{(frac.)}}
\def\qq{$Q^{2}$}
 
\def\wrang{$130 < W < 280\GeV$}
\def\qqrang{\qq${}<1\,\g2$}
\def\ptrang{$3 < \ptdss < 12$\,GeV}
\def\etarang{$-1.5<\eta^{\dss}<1.5$}
\def\xsecdss{3.79\pm0.59\stat \PM{0.26}{0.46} \syst \pm0.94\br\nb}
\def\xsecds{ 9.17\pm0.35\stat \PM{0.40}{0.39}\syst \nb}
\def\xsecra{0.41\pm0.07\stat^{+0.03}_{-0.05}\syst\pm0.10\br} 
\def\gsvalue{0.27\pm0.04\stat \PM{0.02}{0.03}\syst \pm0.01\frcm
  \pm0.07\br}
\def\UA1{0.29\pm0.02\stat\pm{0.01}\syst} 
 
\def\tablesup{\tabcolsep 0pt
  \def\0{\phantom{0}}\def\+{\phantom{-}}
  \def\PM##1##2{\mathop{^{+##1}_{-##2}}}
  \def\E##1##2##3{\pm ##1 \PM{##2}{##3}}
  \def\NL{\\\noalign{\smallskip}}
  \def\RL{\hline\noalign{\smallskip}}
}
\def\Ds{{D_s}}
 
\begin{document} 
 
\title{Measurement of\\ Inclusive \protect\( \dsspm \protect \) 
Photoproduction at HERA}
 
\author{ZEUS Collaboration}
 
\date{}
 
\null\vfil
\begin{quotation}\tt
  \noindent DESY~00-041\hfill ISSN~0418-9833\linebreak
  hep-ex/0003018\\
  March 2000
\end{quotation}
{\let\newpage\relax 
  \maketitle}
\thispagestyle{empty}
 
\begin{abstract}
\noindent The first measurement of inclusive \( \dsspm \)
photoproduction at HERA has been performed with the ZEUS detector for
photon-proton centre-of-mass energies \wrang.  The measured cross
section for
$3 < p_{\!\perp}^\Ds < 12\GeV$ and $|\eta^\Ds|< 1.5$ is
\( \sigma_{ep\to\dss X}= 3.79\pm0.59\stat \PM{0.26}{0.46} \syst
\pm0.94\br\nb \), where the last error arises from the uncertainty in
the $\dsspm$ decay branching ratio.  The measurements are compared
with inclusive \( \dspm \) photoproduction cross sections in the same
kinematic region and with QCD calculations.  The \( \dsspm \) cross
sections lie above a fixed-order next-to-leading order calculation and
agree better with a tree-level \( O(\alpha \alpha _{s}^{3}) \)
calculation that was tuned to describe the ZEUS $\dspm$ cross
sections.  The ratio of \( \dsspm\) to \( \dspm\) cross sections is
$\xsecra$.  From this ratio, the strangeness-suppression factor in
charm photoproduction, within the LUND string fragmentation model, has
been calculated to be $\gs = 0.27\pm 0.05\pm 0.07\br$.  The
cross-section ratio and $\gs$ are in good agreement with those
obtained in charm production in $e^+ e^-$ annihilation.
\end{abstract}
\vfil\vfil\vfil\null
 
\clearpage 
\begin{sloppy}
%
%
%
%
\def\3{\ss} 
\pagenumbering{Roman} 
                                                   %
\begin{center} 
{                      \Large  The ZEUS Collaboration              } 
\end{center} 
  J.~Breitweg, 
  S.~Chekanov, 
  M.~Derrick, 
  D.~Krakauer, 
  S.~Magill, 
  B.~Musgrave, 
  A.~Pellegrino, 
  J.~Repond, 
  R.~Stanek, 
  R.~Yoshida\\ 
 {\it Argonne National Laboratory, Argonne, IL, USA}~$^{p}$
% ...
\newpage 
{\small
% ...
$^{  26}$ supported by a MINERVA Fellowship \\ 
% ...
}                                                           %
                                                           %
\newpage   
                                                           %
                                                           %
\begin{tabular}[h]{rp{14cm}} 
$^{a}$ &  supported by the Natural Sciences and Engineering Research 
          Council of Canada (NSERC)  \\ 
$^{b}$ &  supported by the FCAR of Qu\'ebec, Canada  \\ 
$^{c}$ &  supported by the German Federal Ministry for Education and 
          Science, Research and Technology (BMBF), under contract 
          numbers 057BN19P, 057FR19P, 057HH19P, 057HH29P, 057SI75I \\ 
$^{d}$ &  supported by the MINERVA Gesellschaft f\"ur Forschung GmbH, the 
German Israeli Foundation, the Israel Science Foundation, the Israel 
Ministry of Science and the Benozyio Center for High Energy Physics\\ 
$^{e}$ &  supported by the German-Israeli Foundation, the Israel Science 
          Foundation, the U.S.-Israel Binational Science Foundation, and by 
          the Israel Ministry of Science \\ 
$^{f}$ &  supported by the Italian National Institute for Nuclear Physics 
          (INFN) \\ 
$^{g}$ &  supported by the Japanese Ministry of Education, Science and 
          Culture (the Monbusho) and its grants for Scientific Research \\ 
$^{h}$ &  supported by the Korean Ministry of Education and Korea Science 
          and Engineering Foundation  \\ 
$^{i}$ &  supported by the Netherlands Foundation for Research on 
          Matter (FOM) \\ 
$^{j}$ &  supported by the Polish State Committee for Scientific Research, 
          grant No. 112/E-356/SPUB/DESY/P03/DZ 3/99, 620/E-77/SPUB/DESY/P-03/ 
          DZ 1/99, 2P03B03216, 2P03B04616, 2P03B03517, and by the German 
          Federal Ministry of Education and Science, Research and Technology (BMBF)\\ 
$^{k}$ &  supported by the Polish State Committee for Scientific 
          Research (grant No. 2P03B08614 and 2P03B06116) \\ 
$^{l}$ &  partially supported by the German Federal Ministry for 
          Education and Science, Research and Technology (BMBF)  \\ 
$^{m}$ &  supported by the Fund for Fundamental Research of Russian Ministry 
          for Science and Edu\-cation and by the German Federal Ministry for 
          Education and Science, Research and Technology (BMBF) \\ 
$^{n}$ &  supported by the Spanish Ministry of Education 
          and Science through funds provided by CICYT \\ 
$^{o}$ &  supported by the Particle Physics and 
          Astronomy Research Council \\ 
$^{p}$ &  supported by the US Department of Energy \\ 
$^{q}$ &  supported by the US National Science Foundation 
\end{tabular} 
                                                           %
                                                           %
\end{sloppy}
\clearpage
\pagenumbering{arabic}
 
\section{Introduction}
 
Inclusive \( \dsspm \) photoproduction cross sections at HERA are
presented for photon-proton centre-of-mass energies in the range
\wrang.  The \( \dsspm \) mesons were reconstructed through the decay
chain \( \dsspm \rightarrow \phi \pi ^{\pm }\rightarrow (K^+
K^-)\pi^{\pm} \).  This analysis supplements recent measurements of
inclusive photoproduction of $\dspm$ mesons at HERA~\cite{dstar98,H1}.
The high-statistics measurement by the ZEUS
collaboration~\cite{dstar98} was performed in the same $W$ range as
given above.  The measured cross sections were compared to
next-to-leading order (NLO) calculations~\cite{MassC,LessC.K,LessC.C}
with fragmentation parameters extracted from $\dspm$ production in
$e^+ e^-$ annihilation. The experimental results were found generally
to lie above the NLO expectations, in particular in the forward
(proton) direction.
 
The study of \( \dsspm \) photoproduction provides another test of
perturbative QCD (pQCD) calculations of charm production~\cite{MassC,
  BKL} which is experimentally independent of the \( \dspm \)
measurement. Furthermore, from a ratio of the $\dsspm$ and $\dspm$
cross sections the strangeness-suppression factor, $\gs$, in charm
fragmentation can be determined.  A comparison of the cross-section
ratio and $\gs$ with those obtained in charm production in $e^+ e^-$
annihilation tests the universality of charm fragmentation.

\section{Experimental Conditions}
 
The measurements were performed at the HERA $ep$ collider in the ZEUS
detector during 1996/1997.  In this period HERA collided positrons
with energy $E_e = 27.5\GeV$ and protons with energy $E_p = 820\GeV$.
The integrated luminosity used in this analysis is $38\ipb$.  A
detailed description of the detector can be found
elsewhere~\cite{detector}.
 
Charged particles were measured in the central tracking detector,
CTD~\cite{CTD}, which is a drift chamber consisting of \( 72 \)
concentric sense-wire layers covering the polar angle\footnote{%
  The ZEUS coordinate system is right-handed and has the nominal
  interaction point at \( X=Y=Z=0 \), with the \( Z \)-axis pointing
  in the proton beam direction and the $X$-axis horizontal, pointing
  towards the center of HERA.}  region \( 15\deg <\theta <164\deg \).
The CTD operates in a magnetic field of \( 1.43\, \textrm{T} \)
provided by a thin superconducting solenoid. The transverse momentum
resolution for full length tracks is \( \sigma _{\pt }/\pt
=0.0058\,\pt \oplus 0.0065\oplus {0.0014/\pt } \) (\( \pt \) in GeV).
To estimate the energy loss, \( dE/dx \), of tracks, the truncated
mean of the sense-wire pulse-heights was recorded for each track,
discarding the \( 10\perc \) lowest and \( 30\perc \) highest pulses.
 
The solenoid is surrounded by the uranium\vs scintillator sampling
calorimeter (CAL)~\cite{CAL}, which is almost hermetic and consists of
5918 cells, each read out by two photomultipliers. Under test beam
conditions, the CAL has a relative energy resolution of \(
0.18/\sqrt{\! E} \) (\( E \) in GeV) for electrons and \(
0.35/\sqrt{\! E} \) for hadrons.
 
The luminosity was determined from the rate of the bremsstrahlung
process \( e^{+}p\rightarrow e^{+}\gamma p \), where the photon was
measured by a lead\vs scintillator calorimeter~\cite{Lumi} located at
\( Z=-107\, \textrm{m} \).

\section{\boldmath Event Selection and \protect\( \dsspm \protect \)
  Reconstruction}
 
The ZEUS detector uses a three-level trigger system~\cite{detector}.
At the first level (FLT), the calorimeter cells were combined to
define regional and global sums that were required to exceed various
CAL energy thresholds.  At the second level (SLT), beam-gas events
were rejected by cutting on the quantity $\Sigma_{i}(E-p_{Z})_{i} >
8\GeV$, where the sum runs over all calorimeter cells and $p_{Z}$ is
the $Z$ component of the momentum vector assigned to each cell of
energy $E$.  At the third level (TLT), at least one combination of CTD
tracks was required to be within wide mass windows around the nominal
$\phi$ and
\( \dss \) meson\footnote{%
  Here \( \ds \) and \( \dss \) refer to \( \dspm \) and \( \dsspm \),
  respectively.  } mass values, assuming $\pi$ and $K$ masses as
appropriate.  The $\dss$ transverse momentum, $\ptdss$, was required
to be greater than $2.8\GeV$.
 
Photoproduction events were selected by requiring that no scattered
positron was identified in the CAL~\cite{no-electron}.  The Jacquet\vs
Blondel~\cite{JB} estimator of $W$, $\WJB = \sqrt{2 E_{p}
  {\Sigma_{i}(E-p_{Z})_{i}}}$, was required to be between 115 and 250
GeV. The lower limit was due to the SLT requirement, while the upper
one suppressed remaining DIS events with an unidentified scattered
positron in the CAL~\cite{no-electron}.  After correcting for detector
effects, the most important of which were energy losses in inactive
material in front of the CAL and particle losses in the beam
pipe~\cite{no-electron,dstar95}, this $\WJB$ range corresponds to an
interval of true~\( W \) of \wrang.  Under these conditions, the
photon virtuality, \( Q^{2} \), is limited to values less than \( 1
\g2 \).  The corresponding median \( Q^{2} \) was estimated from a
Monte Carlo (MC) simulation to be about \( 3 \times 10^{-4}\g2 \).
 
The MC sample used for this analysis was prepared with the PYTHIA
6.1~\cite{PYTHIA} generator. The proportions of direct- and
resolved-photon events~\cite{dirres} corresponded to the PYTHIA cross
sections ($\approx 50\%$ each), where charm excitation processes were
included in the resolved component~\cite{dstar98}.  The
MRSG~\cite{MRSG} and GRV-G~HO~\cite{GRV} parametrisations were used
for the proton and photon structure functions, respectively. The MC
events were processed through the standard ZEUS detector- and
trigger-simulation programs and through the event reconstruction
package used for offline data processing.  The data and MC
distributions were found to be in good agreement.
 
The \( \dss \) mesons were reconstructed through the decay mode \(
\dsspm \rightarrow \phi \pi ^{\pm } \), which has a branching ratio \(
B_{\dss \rightarrow \phi \pi }=0.036\pm 0.009 \)~\cite{PDG}.  The \(
\phi \) was identified via its decay mode \( \phi \rightarrow
K^{+}K^{-} \), with \( B_{\phi \rightarrow K^+K^- }= 0.491\pm 0.008
\)~\cite{PDG}.  The analysis was restricted to the pseudorapidity
range \( -1.5<\eta ^{\dss }<1.5 \), for which the CTD acceptance is
high.  Here \( \eta ^{\dss }\equiv -\ln (\tan \frac{\theta }{2}) \),
where \( \theta \) is the polar angle with respect to the proton beam
direction.  The kinematic region in $\ptdss$ was limited to \(
3<\ptdss <12\GeV \). The lower cut was required to comply with the
$\ptdss$ cut applied at the TLT\@.  The upper cut was due to the limited
statistics.
 
The three \( \dss \) decay tracks were required to originate from the
event vertex, which was measured with a resolution of 0.4 cm in the
$Z$ direction and 0.1 cm in the $XY$ plane. Only tracks with polar
angles \( 20^{\circ }<\theta <160^{\circ } \) and transverse momenta
\( \pt >0.75\, \GeV \) were considered in the analysis.
 
The decay of the pseudoscalar \( \dss \) meson to the $\phi$ (vector)
plus $\pi$ (pseudoscalar) final state results in an alignment of the
spin of the $\phi$ meson with respect to the direction of motion of
the $\phi$ relative to the \( \dss \). Consequently, the distribution
of \( \cos \theta ^{*}_{K} \), where \( \theta ^{*}_{K} \) is the
angle between one of the kaons and the pion in the \( \phi \)~rest
frame, followed a \( \cos ^{2}\theta ^{*}_{K} \) shape, implying a
flat distribution for \( \cos ^{3}\theta ^{*}_{K} \).  In contrast,
the \( \cos \theta ^{*}_{K} \) distribution of the combinatorial
background was flat and its \( \cos ^{3}\theta ^{*}_{K} \)
distribution peaked at zero.  A cut of \( |\cos^{3}\theta
^{*}_{K}|>0.15 \) suppressed the background by a factor of
approximately two while reducing the signal by 15\%.
 
\begin{figure}
  {\centering {\Large\sffamily ZEUS 96+97\par}\medskip
  \noindent\parbox[t]{0.65\hsize}
  {\vglue0pt\resizebox*{\hsize}{!}
{\includegraphics{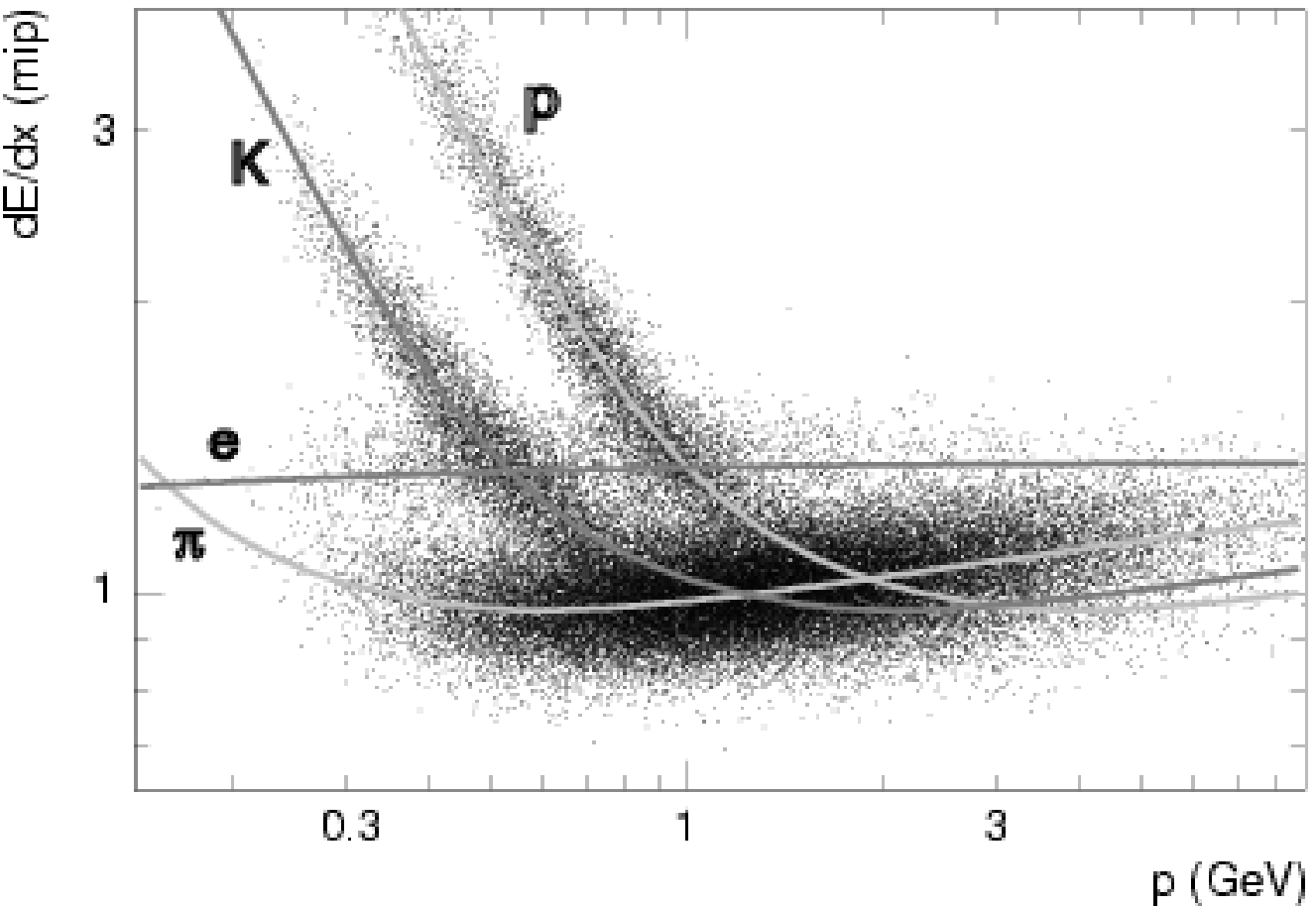}}}
\llap{\parbox[t]{7mm}{\vglue3mm\sffamily(a)}}\par} \smallskip
\hbox to \hsize{%
  \parbox[t]{0.49\hsize}{\vglue0pt\resizebox*{\hsize}{!}
  {\includegraphics{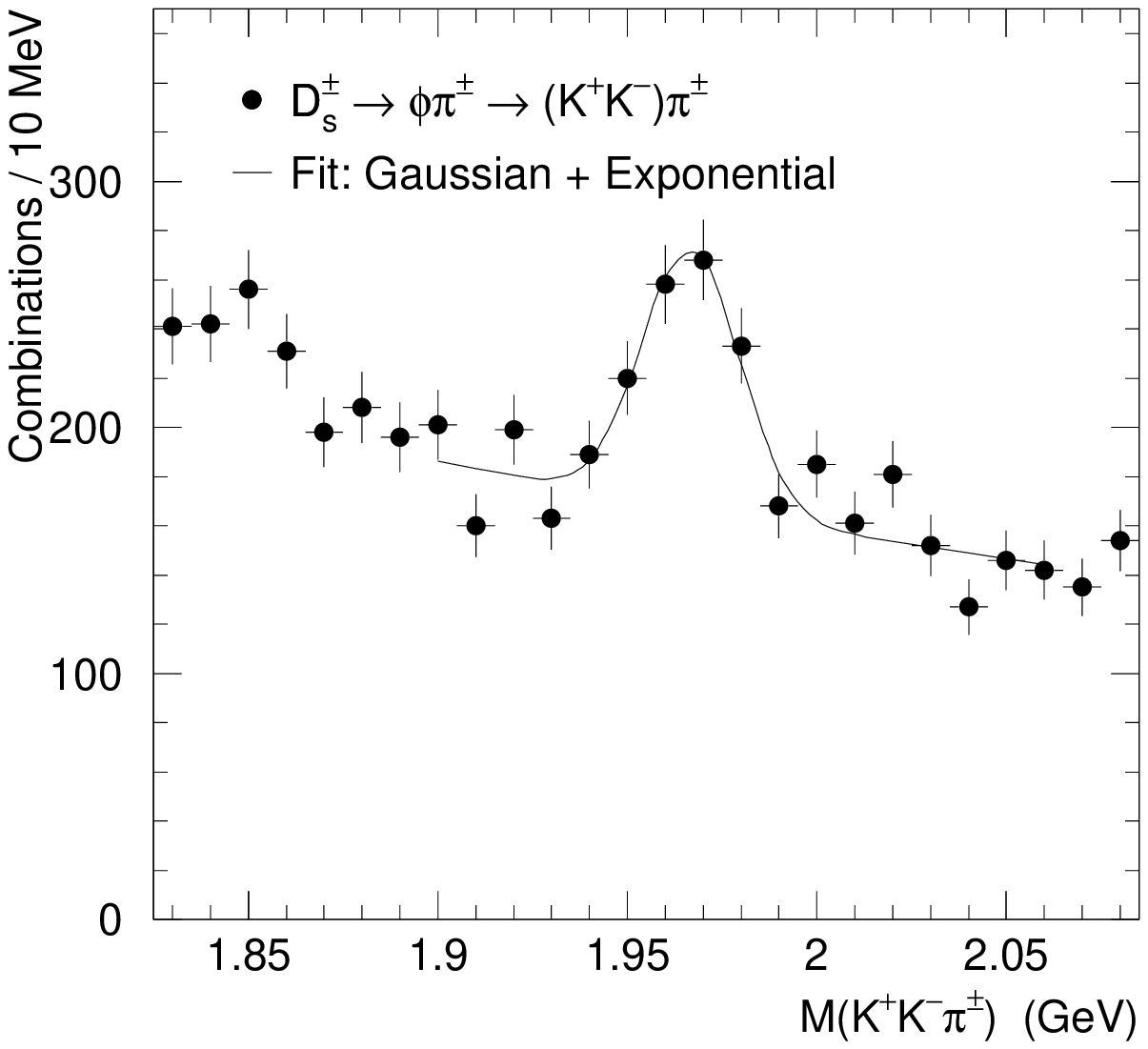}}}%
\llap{\parbox[t]{7mm}{\vglue3mm\sffamily(b)}}\hfill
\parbox[t]{0.49\hsize}{\vglue0pt\resizebox*{\hsize}{!}
  {\includegraphics{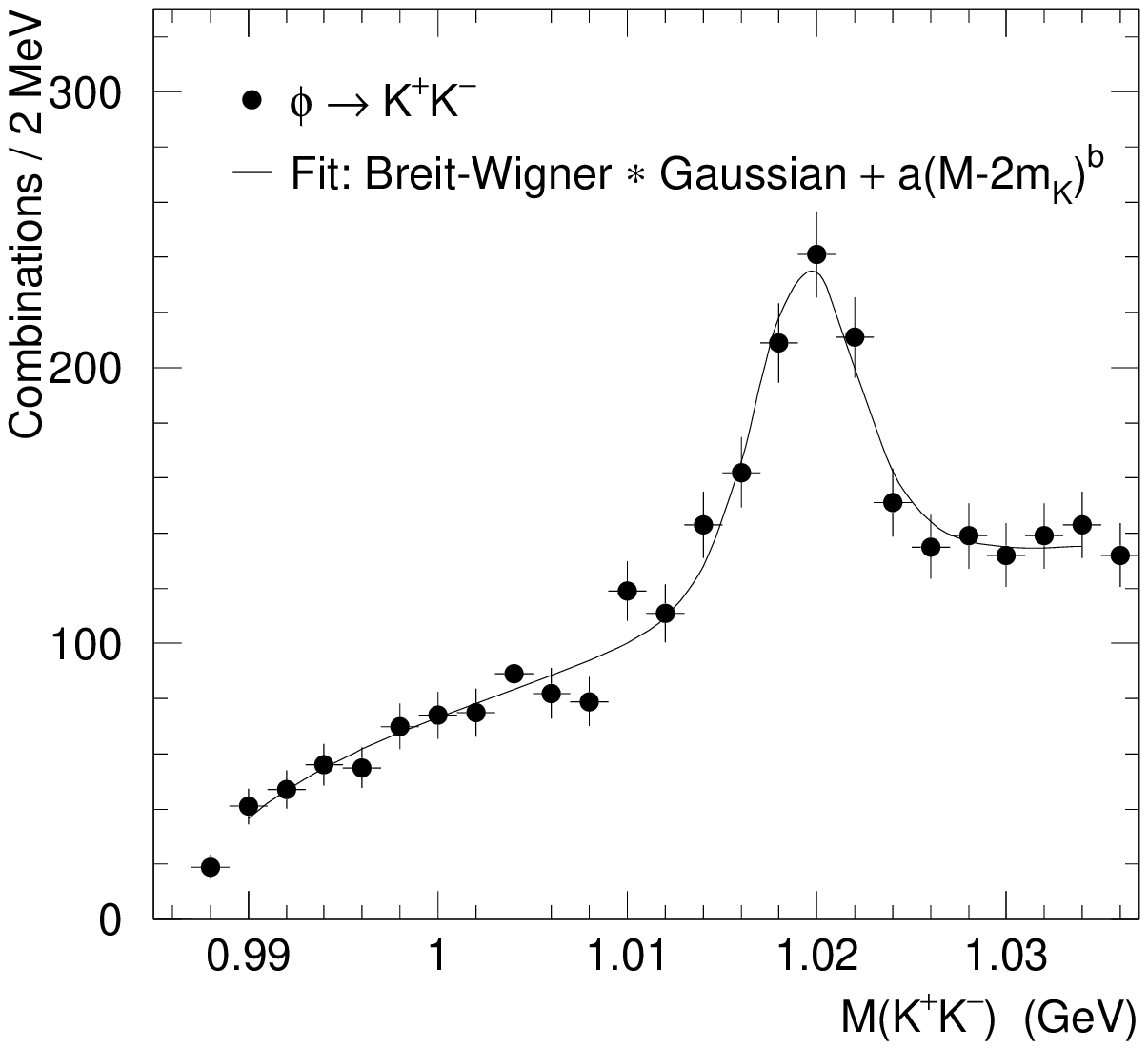}}}
\llap{\parbox[t]{9mm}{\vglue3mm\sffamily(c)}}}
\caption[]{(a)~Distribution of the sample used for fitting the
  $dE/dx$~parametrisation in the $(dE/dx,p)$~plane. Here $dE/dx$
  is normalised to a minimum ionizing particle (MIP), defined as the
  average truncated mean of pion tracks in the momentum range \(
  0.3<p<0.4\GeV \).  The lines indicate the fit
  result~\cite{olaf}.\enspace (b)~\( M(K^+ K^- \pi^{\pm} ) \)
  distribution for events inside the \protect\( \phi \protect \) mass
  range (\protect\( 1.0115<M(K^{+}K^{-})<1.0275\GeV \protect \)).  The
  solid curve is a fit to a Gaussian (representing the resonance) plus
  an exponential background.\enspace (c)~\protect\( M(K^{+}K^{-})\protect \)
  distribution for events inside the \protect\( \dss \protect \) mass
  range (\protect\( 1.94<M(K^{+}K^{-}\pi^{\pm} )<2.00\GeV \protect
  \)).  The solid curve is a fit to a Breit-Wigner convoluted with a
  Gaussian-shaped resonance and a background parametrisation of the
  form \mbox{$a [ M(K^{+}K^{-}) - 2 m_K]^{b}$}.}
\label{fig:dss-signal}
\end{figure}

The \( dE/dx \) information was used to allow partial \( K \) and \(
\pi \) separation.  For each kaon (pion) candidate a likelihood
function \( \ell _{K(\pi )}\equiv \exp
\{-\frac{1}{2}[(dE/dx)_{\textrm{meas}}-(dE/dx)_{K(\pi )}]^{2}/\sigma
^{2}\} \) was determined, where $(dE/dx)_{K(\pi )}$ is the expected
value for a kaon (pion), and \( \sigma \) is the $dE/dx$ resolution,
which is inversely proportional to $\sqrt{n}$, where $n$ is the number
of hits entering the truncated mean.  The parametrisation for the
$dE/dx$~expectation was obtained~\cite{olaf} from a fit to an
independent inclusive track sample (Fig.~\ref{fig:dss-signal}a).  A
normalised likelihood function \( L_{i}\equiv \ell _{i}/\sum_j \ell
_{j} \) was defined, where the sum extends over the considered
particle hypotheses $\pi$, $K$ and $p$.  Provided that the number of
hits was sufficiently high (\( n>7 \)), low-likelihood hypotheses were
rejected if \( \ell _{i}<0.05 \), unless \( L_{i}>0.12 \).  The
$dE/dx$ cuts reduced the combinatorial background by approximately
20\%.  The signal loss was determined by means of a Monte Carlo
simulation, using $dE/dx$ parameters obtained from the data.  The
overall loss was $2.1\%$.
 
For the \( \dss \) candidates, $p_{\perp}^{\dss} / E_{\perp}^{\theta >
  10^{\circ}} >0.18$ was required, where \( E_{\perp}^{\theta >
  10^{\circ}} \) is the transverse energy outside a cone of \( \theta
=10^{\circ } \) defined with respect to the proton direction. This cut
removed more than $20\%$ of the background while preserving about
$95\%$ of the \( \dss \) signal, as verified by MC studies.
 
The \( K^+ K^- \pi^{\pm} \) mass distribution with the above cuts is
shown in Fig.~\ref{fig:dss-signal}b for events in the $\phi$ mass
range, \protect\( 1.0115<M(K^{+}K^{-})<1.0275\GeV \protect \).  The
fraction of events with more than one entry in the $\dss$ mass region
was less than 1\%.  The mass distribution was fitted to a sum of a
Gaussian with the $\dss$ mass and width as free parameters, and an
exponential function describing the non-resonant background. In order
to avoid a possible contribution from $D^{\pm}\to\phi\pi^{\pm}$, the
fit was not extended below $1.895\GeV$.  The fit yielded \( 339\pm48
\) \( \dss \)~mesons.  The mass value obtained was $M_{\dss}=1967\pm
2\MeV$, in agreement with the PDG value~\cite{PDG}. The width of the
signal was $\sigma_{\dss}=12.5\pm 1.9\MeV$, in agreement with the MC
estimation.
 
A clear $\phi$ signal is seen in the \( M(K^{+}K^{-}) \) distribution
(Fig.~\ref{fig:dss-signal}c) for the \( \dss \) region, \(
1.94<M(K^{+}K^{-}\pi^{\pm} )<2.00\GeV \).  The fit function for the
$\phi$ was a relativistic P-wave Breit-Wigner with variable mass and a
fixed full-width of $4.43\MeV$~\cite{PDG}, convoluted with a Gaussian
function whose width, $\sigma_{\phi}$, was a free parameter of the
fit. The background was parametrised with the functional form \mbox{$a
  [ M(K^{+}K^{-}) - 2 m_K]^{b}$}, where $m_K$ is the $K^{\pm}$ mass.
The fit yielded $M(\phi)= 1019.5\pm 0.3\MeV$, in agreement with the
PDG value~\cite{PDG}, and $\sigma_{\phi}=1.7\pm 0.4\MeV$, in agreement
with MC estimation.  The number of $\phi$ mesons originating from
$\dss$ decays was estimated by a side-band subtraction, and was found
to be in good agreement with the number of $\dss$ obtained from the
above $M(K^+ K^- \pi^{\pm})$ fit.

\section{\boldmath Measurement of Inclusive $\dss$ Cross Sections}
 
The inclusive $\dss$ cross section is given by:
\[ \sigma_{ep\to\dss X}=\frac{N_{\dss}}{A\cdot\mathcal{L}\cdot B_{\dss
    \rightarrow (\phi {\rightarrow} K^+ K^-)\pi }}\;,
\]
where $N_{\dss}$ is the fitted number of $\dss$ mesons, $\mathcal{L}$
is the integrated luminosity, \( B_{{\scriptscriptstyle \dss \to (\phi
    \to K^+ K^-)\pi }}\linebreak =0.0177\pm 0.0044 \) is the combined
\( \dss \rightarrow (\phi {\to} K^+ K^-)\,\pi \) decay branching ratio
and $A$ is the acceptance calculated with the MC sample (Section 3).
The MC sample contains all events with the $\dss\to(\phi{\to} K^+
K^-)\pi$ decay channel as well as small admixtures from other $\dss$
decay modes and from other charm particle decays. Thus these
contributions were taken into account in the acceptance correction
procedure.
 
\begin{figure}
  {\centering \parbox[t]{0.57\hsize}{\vglue0pt\resizebox*{\hsize}{!}
  {\includegraphics{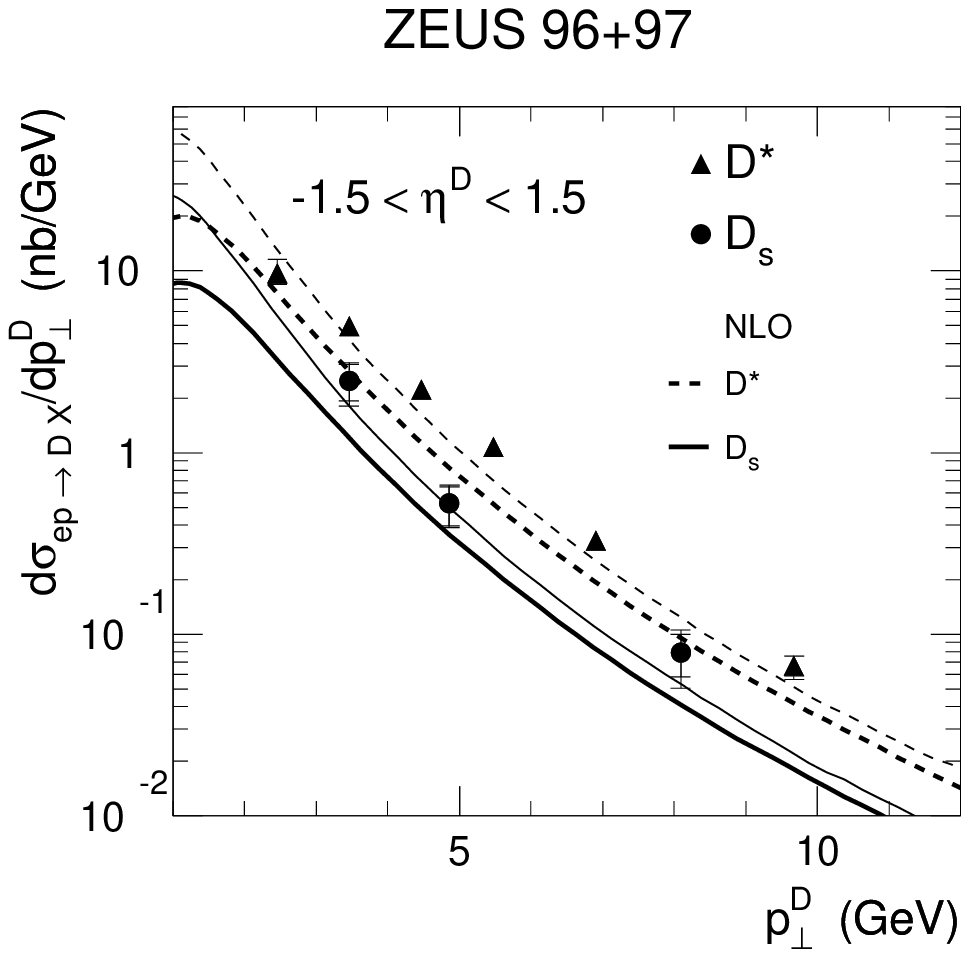}}}%
\llap{\parbox[t]{10mm}{\vglue13mm\sffamily(a)}}\par\bigskip
\parbox[t]{0.57\hsize}{\vglue0pt\resizebox*{\hsize}{!}
  {\includegraphics{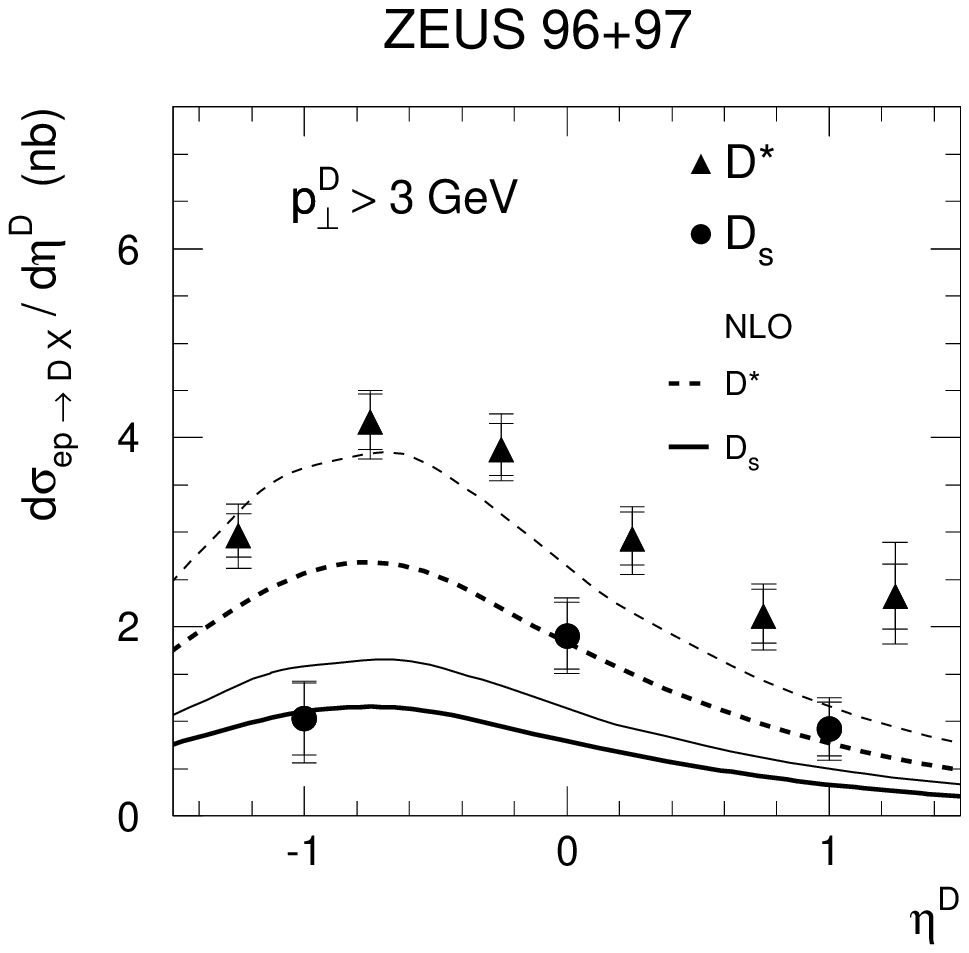}}}%
\llap{\parbox[t]{10mm}{\vglue13mm\sffamily(b)}}\par}
\caption[]{Differential cross sections for the photoproduction
  reaction \( ep\to D\, X \): (a)~${d\sigma} / {d\pt ^{D}}$ and
  (b)~${d\sigma }/{d\eta ^{D}}$, where \protect\( D\protect \) stands
  for \protect\( \ds \protect \) or \protect\( \dss \protect \). The
  $\pt ^{D}$ points are drawn at the position of the average value of
  an exponential fit in each bin. The $\eta ^{D}$ points are drawn at
  the middle of each bin.  The inner error bars show the statistical
  uncertainty, while the outer ones show the statistical and
  systematic errors added in quadrature. Normalisation uncertainties
  due to the \( \dss \rightarrow \phi \pi \) branching ratio are not
  included in the systematic errors or in the theoretical
  calculations.  The \protect\( \dss \protect \) (dots) and \protect\(
  \ds \protect \) (triangles) data are compared with the NLO
  predictions for $\dss$ (full curves) and $\ds$ (dashed curves) with
  two parameter settings: \( m_{c}=1.5\GeV \), \( \mu _{R} = m_{\perp
    } \) (thick curves) and \( m_{c}=1.2\GeV \), \( \mu _{R} =
  0.5m_{\perp } \) (thin curves).}
\label{fig:dss-nlo}
\end{figure}

The total \( \dss \) cross section in the kinematic region \qqrang,
\wrang, \ptrang\ and \etarang\ was measured to be \( \sigma_{ep\to\dss
  X}=\xsecdss \), where the last error is due to the 25\% uncertainty
in the branching ratio \( B_{\dss \rightarrow \phi \pi } \).  The
differential cross sections \( d\sigma /d\ptdss \) and \( d\sigma
/d\eta ^{\dss } \) are given in Table 1.
\begin{table}\tablesup
 \caption[]{The differential cross sections $d\sigma/dp_{\!\perp}^\Ds$ 
  for $|\eta^\Ds|< 1.5$ and $d\sigma/{d\eta^\Ds}$ for $3 <
  p_{\!\perp}^\Ds < 12\GeV$ for the photoproduction reaction \(
  ep\to\dss X \) in various $p_{\!\perp}^\Ds$ and $\eta^\Ds$ bins
  for \qqrang\ and \wrang. The $p_{\!\perp}^\Ds$ points are given at
  the positions of the average values of an exponential fit to the
  $p_{\!\perp}^\Ds$ distribution in each bin. Statistical and
  systematic uncertainties are quoted separately; the third error
  corresponds to the uncertainty in the $\dss \to \phi\pi$ branching
  ratio.}
 \def\bigbran{\smash{\left.\begin{matrix}\strut\\\strut\\\strut\end{matrix}
 \right\rbrace}\pm25\%}
 \begin{tabular*}{\hsize}{c@{ -- }cl@{\tabskip\fill}lll@{\tabskip0pt}l}\RL%
  \multicolumn3l{$p_{\!\perp}^\Ds$ range [GeV]}& 
  ${\langle p_{\!\perp}^\Ds\rangle}_{\text{fit}}$ [GeV]& 
  $d\sigma/dp_{\!\perp}^\Ds
 \pm\text{stat.}\pm\text{syst.}\pm\text{br.}$ [nb/GeV]\NL\RL
  3&\04&& 3.46 &$ 2.49\0\E{0.56\0}{0.32\0}{0.40\0}$\NL
  4&\06&& 4.86 &$ 0.523 \E{0.127} {0.058} {0.052}\bigbran$\NL
  6& 12&& 8.09 &$ 0.079 \E{0.021} {0.017} {0.020}$\NL
  \RL
  \multicolumn3l{$\eta^\Ds$ range}& &
  $d\sigma/d\eta^\Ds
 \pm\text{stat.}\pm\text{syst.}\pm\text{br.}$ [nb]\NL\RL
 -1.5&{-0.5}&&&$ 1.03\0\E{0.38\0}{0.09}{0.26}$\NL
 -0.5&\+0.5 &&&$ 1.90\0\E{0.35\0}{0.20\0}{0.19}\bigbran$\NL
  \+0.5&\+1.5 &&&$ 0.92\0\E{0.29\0}{0.16}{0.17}$\NL
  \RL
 \end{tabular*}
\end{table}
In Fig.~\ref{fig:dss-nlo} they are compared with the distributions for
the \( \ds \) in the same kinematic region. The overall normalisation
uncertainty due to the luminosity measurement (1.7\%) is not included
in the cross section errors.

\subsection{Systematic Uncertainties}
 
A detailed study of possible sources of systematic uncertainties was
carried out for all measured cross sections by shifting the nominal
analysis parameters as described below, taking into account resolution
effects.  For each variation, except for the first one, which is due
to the fit systematics, the $\dss$ mass and width were fixed to the
values found for the nominal cut values.  The following sources of
systematic errors were considered:
 
\begin{itemize}
\item the uncertainties in the determination of the number of $\dss$
  mesons were estimated by using a quadratic polynomial function for
  the background parametrisation instead of an exponential one and by
  varying the range of the \( K^+ K^- \pi^{\pm} \) mass distribution
  in the fit procedure;
\item to estimate the uncertainties in the tracking procedure, the
  track selection cuts, including $M(K^+ K^- )$ and \( \cos \theta
  ^{*}_{K} \), were shifted by at least the expected resolutions from
  the nominal values (Section~3);
\item the cut on $p_{\perp}^{\dss} / E_{\perp}^{\theta > 10^{\circ}}$
  was changed by $\pm 10\%$;
\item the MC simulation was found to reproduce the absolute energy
  scale of the CAL to within $\pm 3\%$~\cite{Escale}. The
  corresponding uncertainty in the cross section was calculated,
  including a shift of 3\% in the FLT CAL energy thresholds (Section
  3);
\item the fraction of resolved photon processes in the PYTHIA MC
  sample was varied between 40\% and 60\%;
\item the $dE/dx$ likelihood cuts were changed in the range $0.02 <
  l_i < 0.10$ and $0.08 < L_i < 0.15$;
\item the uncertainty associated with the correction to the true $W$
  range was determined by moving the $\WJB$ boundary values by the
  estimated resolution of $\pm 7\%$.
\end{itemize}
 
None of the above was dominant in the total systematic uncertainty to
the inclusive $\dss$ cross section. All systematic errors were added
in quadrature, yielding a total uncertainty of $^{+7}_{-12}\%$,
compared to a statistical error of 16\%.

\section{\boldmath Comparison with QCD Calculations } 
 
The $\dss$ cross sections were compared to two types of pQCD
calculations~\cite{MassC, BKL}. The fractions of \( c \) quarks
hadronising as \( \ds \) and \( \dss \) mesons were used as input to
each calculation. The values \( f(c\rightarrow D^{*+})=0.235\pm 0.007
\mathop{(\pm} 0.007) \) and \( f(c\rightarrow D_s^{+})=0.101\pm 0.009
\mathop{(\pm} 0.025) \) were extracted~\cite{LG} by a least-squares
procedure from all relevant existing $e^+ e^-$ experimental
data~\cite{epem}. The errors in brackets are due to uncertainties in
the charm hadron decay branching ratios.  They affect experimental and
theoretical cross-section calculations in the same way and can be
ignored in the comparison.
 
The NLO calculation of charm photoproduction in the fixed-order
approach of Frixione et al.~\cite{MassC} assumes that gluons and light
quarks $(u,d,s)$ are the only active partons in the structure
functions of the proton and the photon. In this approach there is no
explicit charm excitation component, which can be important in charm
photoproduction at HERA~\cite{dstar98}, and charm is only produced
dynamically in hard pQCD processes. This calculation is expected to be
valid when the $c$-quark transverse momentum, $p_{\perp}$, is not much
larger than the $c$-quark mass, $m_c$.
 
The structure function parametrisations used in the NLO calculation
were MRSG~\cite{MRSG} for the proton and GRV-G HO~\cite{GRV} for the
photon.  The renormalisation scale used was $\mu _{R} =m_{\perp
  }\equiv \sqrt{m_{c}^{2}+\pt ^{2}} $, and the factorisation scales of
the photon and proton structure functions were set to $\mu_{F} = 2
m_{\perp}$. The pole mass definition is used in this calculation for
the $c$-quark mass with a nominal value $m_{c} =1.5\GeV$.  The
Peterson fragmentation function~\cite{Peterson} was used for charm
fragmentation in this calculation.  The Peterson parameter \( \epsilon
= 0.035 \) was obtained for $\ds$ in a NLO fit~\cite{nason99} to ARGUS
data~\cite{epem}. A recent NLO fit~\cite{oleari} to ARGUS data yields
an \( \epsilon (\dss ) \) value equal to \( \epsilon (\ds ) \) within
the fit uncertainties. Using the same value for both channels leads to
the same predictions, except for the difference in \( f(c\rightarrow
D^{*+}) \) and \( f(c\rightarrow D_s^{+}) \), which enter the
calculation as scale factors. The NLO prediction for the total
inclusive $\dss$ cross section of \( 2.18\nb \) is smaller by $\approx
2$ standard deviations compared to the measured cross section.
Scaling the Peterson parameter $\epsilon$ with the squared ratio of
the constituent quark masses~\cite{Peterson}, $m_{s}=0.5\GeV$ and
$m_{u,d}=0.32\GeV$~\cite{jetset}, leads to \( \epsilon (\dss )=0.085
\), which yields a NLO prediction 22\% lower than that with \(
\epsilon (\dss )=0.035 \).
 
In Fig.~\ref{fig:dss-nlo}, the NLO calculations are compared to the
differential cross sections.  The thick curves correspond to the
nominal values of $\mu _{R}$ and $m_{c}$, as defined above. For the
thin curves, a rather extreme value for the $c$-quark mass, \(
m_{c}=1.2\GeV \), and a \( \mu _{R} \) value of \( 0.5m_{\perp } \)
have been used. The \( \dss \) cross section decreases steeply with
rising \( \pt \), as in the \( \ds \) case. The NLO calculation
reproduces within errors the shape of the \( \ptdss \)~distribution
but underestimates the data for the nominal parameter set. For the \(
\eta^{\dss} \)~distribution the NLO predictions are below the data in
the central and forward regions. A similar effect was observed when
${d\sigma }/{d\eta ^{\ds }}$ distributions were compared with various
NLO predictions over a wide range of $W$ values and photon
virtualities~\cite{dstar98, eps99, dsdis}.
 
Recently Berezhnoy, Kiselev and Likhoded (BKL) have suggested another
model~\cite{BKL} which does not employ any specific fragmentation
function. In this tree-level pQCD \( O(\alpha \alpha _{s}^{3}) \)
calculation, the \( (c,\bar{q}) \) state produced in pQCD is
hadronised, taking into account both colour-singlet and colour-octet
contributions. Using the experimental value for \( f(c\rightarrow
D^{*+}) \), a $c$-quark mass $m_c=1.5\GeV$, a light constituent quark
mass $m_q=0.3\GeV$ and the CTEQ4M proton structure function
parametrisation~\cite{cteq}, the ratio of the colour-octet and
colour-singlet components was tuned to describe the ZEUS \(
\ds \) photoproduction cross sections~\cite{dstar98}.\footnote{%
A comparison of the $\ds$ data with the BKL calculations can be found in the   
BKL paper~\cite{BKL}.}
                                                       This ratio and
the experimental value for \( f(c\rightarrow D_s^{+}) \) were used to
obtain predictions for $\dss$ photoproduction. In this case, a strange
quark mass, $m_s=0.5\GeV$, was used instead of $m_q$. The calculated
cross sections for $\dss$ and $\dss^{*\pm}$ were combined, since the
inclusive $\dss$ channel includes fully the prompt $\dss^{*\pm}$ meson
production.
 
\begin{figure}
  {\centering \parbox[t]{0.59\hsize}{\vglue0pt\resizebox*{\hsize}{!}
  {\includegraphics{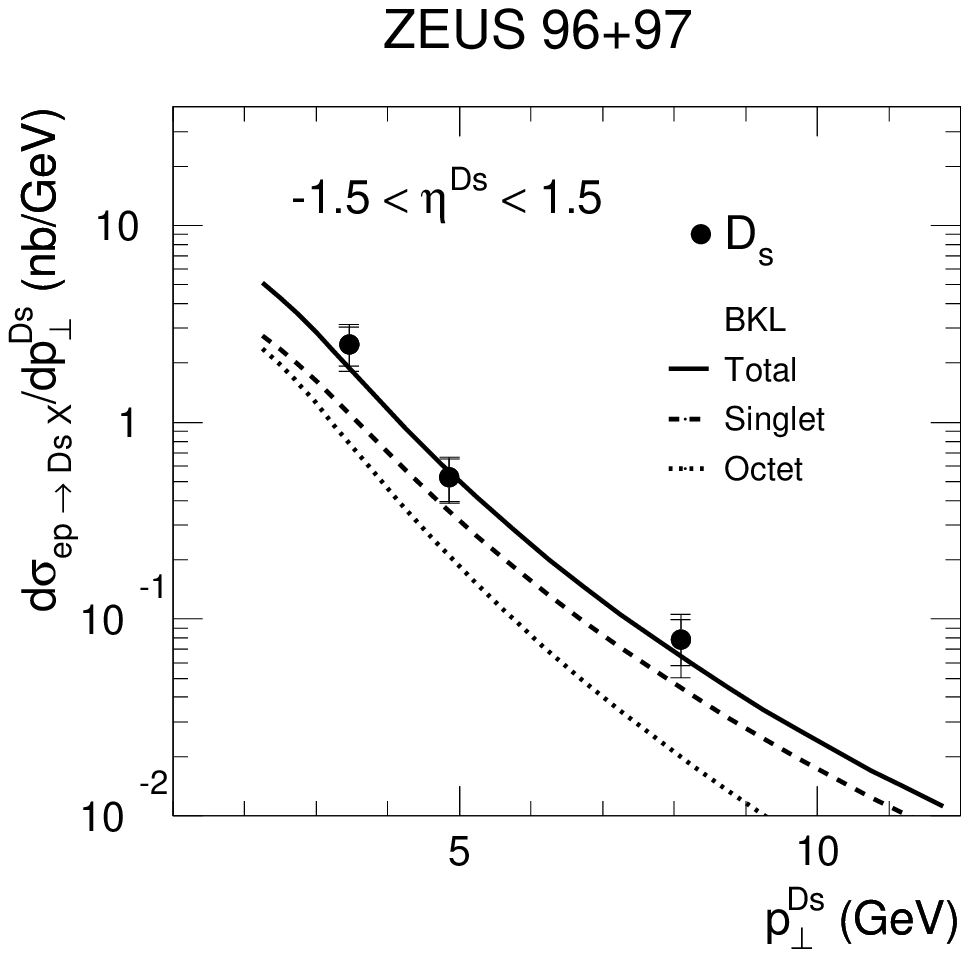}}}%
\llap{\parbox[t]{10mm}{\vglue13mm\sffamily(a)}}\par\bigskip
\parbox[t]{0.59\hsize}{\vglue0pt\resizebox*{\hsize}{!}
  {\includegraphics{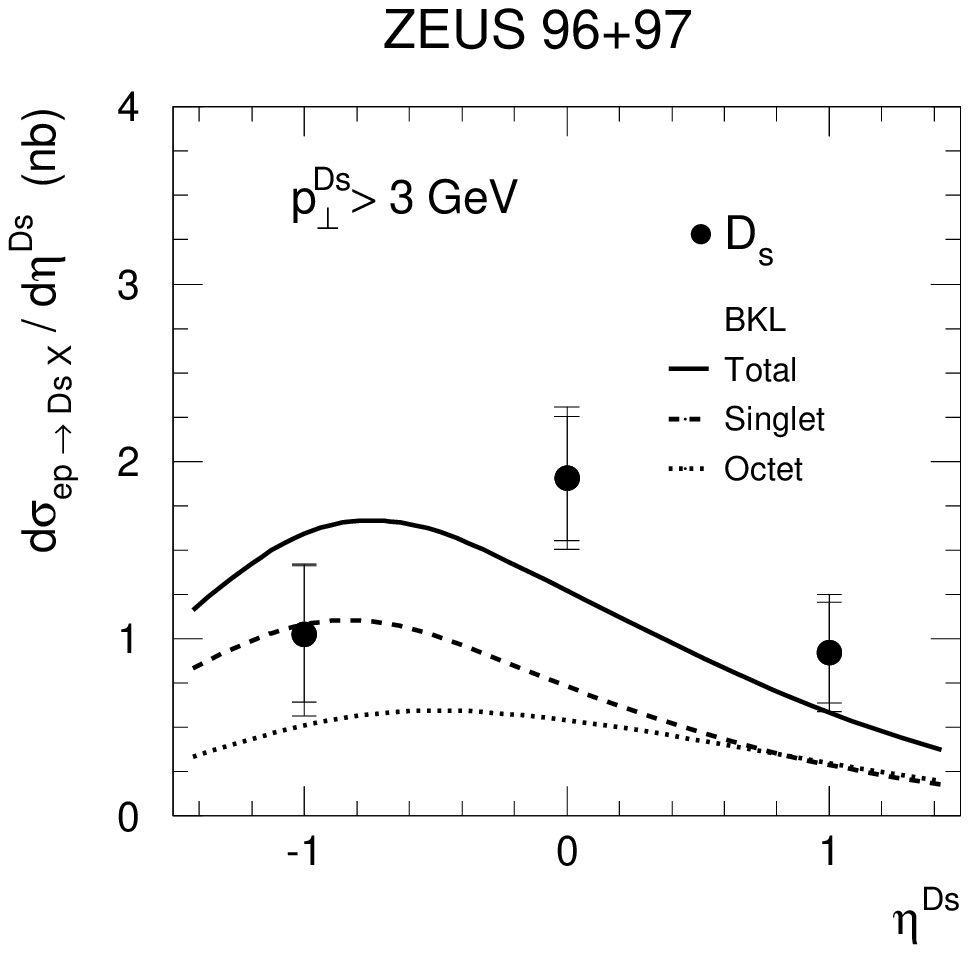}}}%
\llap{\parbox[t]{10mm}{\vglue13mm\sffamily(b)}}\par}
\caption[]{Differential cross sections for the photoproduction
  reaction \( ep\to \dss X \): (a)~${d\sigma} / {d\ptdss }$ and
  (b)~${d\sigma }/{d\eta ^{\dss }}$ compared to the BKL
  model~\cite{BKL}. Colour-singlet (dashed curves) and colour-octet
  (dotted curves) contributions are plotted separately. Their sum is
  shown as the full curves. The $\ptdss$ points are drawn at the
  position of the average value of an exponential fit in each bin.
  The $\eta^{\dss}$ points are drawn at the middle of each bin. The
  inner error bars show the statistical uncertainty, while the outer
  ones show the statistical and systematic errors added in quadrature.
  Normalisation uncertainties due to the \( \dss \rightarrow \phi \pi
  \) branching ratio are not included in the systematic errors or in
  the theoretical calculations.}
  \label{fig:dss-bkl}
\end{figure}

In Fig.~\ref{fig:dss-bkl},
the BKL calculations are compared to the $\dss$ differential
cross-section measurements.  The agreement with the data is better
than that of the NLO calculation with the nominal parameters, but the
shape of the \( \eta ^{\dss } \) distribution is not matched by the
BKL prediction. The total predicted $\dss$ cross section in the
kinematic range of the measurement (\( 3.37\nb \)) is compatible with
the measured inclusive cross section.

\section{\boldmath $\dss$ to $\ds$ cross-section ratio and 
  \protect\( \gs \protect \)}
 
In the $\dss$ kinematic region, as defined in Section 4, the \( \ds
\) cross section was measured to be \( \sigma_{ep\to\ds X}= \xsecds
\)~\cite{dstar98}. This yields a ratio \( \sigma_{ep\to\dss X}/
\sigma_{ep\to\ds X} =\xsecra \), where common systematic errors
($W_{JB}$ and CAL energy scale) have been removed and the last error
is the uncertainty in \( B_{\dss \rightarrow \phi \pi } \). This ratio
is in good agreement with the ratio \( f(c\rightarrow D_s^{+}) /
f(c\rightarrow D^{*+}) = 0.43\pm 0.04\pm 0.11\br \) obtained from
results of $e^+ e^-$ experiments (see Section 5). The last error
originates from the uncertainty in \( B_{\dss \rightarrow \phi \pi }
\) and can be ignored in the comparison.
 
The strangeness-suppression factor, $\gs$, is the ratio of
probabilities to create \( s \) and \( u,d \) quarks during the
fragmentation process.  In simulation programs based on the Lund
string fragmentation scheme~\cite{jetset}, \( \gs \) is a free
parameter with a default value of $0.3$. By varying the value of \(
\gs \) in the PYTHIA simulation~\cite{PYTHIA}, a direct relation can
be obtained between \( \gs \) and the $\dss$ to $\ds$ cross-section
ratio. Fixing the value of $f(c\rightarrow D^{*+})$ in PYTHIA to
0.235~\cite{LG} yields
\[ \gs =\gsvalue\; . \]
The third error is due to the uncertainty in \( f(c\rightarrow D^{*+})
\), while the forth one results from the uncertainty in \( B_{\dss
  \rightarrow \phi \pi } \). Adding all errors in quadrature, except
the last one, gives $\gs = 0.27\pm 0.05\pm 0.07\br$.
 
Previously, $\gs$ was measured mainly from the ratio of $K$ to $\pi$
production and from the momentum spectrum of $K$ mesons in
hadron-hadron~\cite{comp,UA1} and $e^+ e^-$~\cite{Z0,SLD,OPAL}
collisions, as well as in deep inelastic scattering (DIS)
experiments~\cite{neutrino,E665,hera}. The most accurate measurement,
obtained in $p\bar p$ collisions~\cite{UA1}, is $\gs = \UA1$.  The DIS
results require a lower value, $\gs \approx 0.2$. Recent results from
$e^+ e^-$ collisions are in some disagreement with each other. The SLD
preliminary result~\cite{SLD} is $\gs =0.26\pm 0.06$, while OPAL
finds~\cite{OPAL} $\gs =0.422\pm 0.049\pm 0.059$. Previous $\gs$
measurements are therefore in good agreement with that of this
analysis, except the latest OPAL value, which is about 2 standard
deviations higher.
 
Existing $\gs$ values obtained from heavy-meson production (charm and
beauty) in $e^+ e^-$ collisions~\cite{Z0} are centred around 0.3.  For
charm production, the ratio \( 2 f(c\rightarrow D_s^{+}) /
[f(c\rightarrow D^{+}) + f(c\rightarrow D^{0})] \) was used as a
measure of $\gs$.  Using for the above fractions the more recent
values quoted in~\cite{LG} leads to $\gs = 0.26\pm 0.03\pm 0.07\br$,
where the latter uncertainty originates from \( B_{\dss \rightarrow
  \phi \pi } \) and can be ignored in the comparison with the ZEUS
result.
 
The results presented here on the $\dss$ to $\ds$ cross-section ratio
and on $\gs$, taken together with charm production data in $e^+ e^-$
annihilation, tend to support the universality of charm fragmentation.

\section{Summary and Conclusions}
 
The first measurement at HERA of inclusive \( \dss ^{\pm } \)
photoproduction has been performed with the ZEUS detector. The cross
section for \( Q^{2}<1\g2 \), \wrang, \( 3<\ptdss <12\GeV \) and \(
-1.5<\eta^{\dss} <1.5 \) is \( \sigma_{ep\to\dss X}=\xsecdss \). The
differential cross sections \( d\sigma /d\ptdss \) and \( d\sigma
/d\eta ^{\dss } \) are generally above fixed-order NLO calculations,
as was the case with the results previously obtained for \( \ds \)
photoproduction in the same kinematic region. The BKL calculation,
using the octet-to-singlet ratio tuned to fit the ZEUS \( \ds \) data,
describes the $\dss$ cross sections reasonably well, but the shape of
the \( \eta ^{\dss } \) distribution is not matched by the BKL
prediction. The cross-section ratio, \( \sigma_{ep\to\dss X}/
\sigma_{ep\to\ds X} \), in the kinematic region as defined above is
$\xsecra$, in good agreement with the ratio of $c$ quarks hadronising
into $\dss$ and $\ds$ mesons, extracted from $e^+ e^-$ experiments.
From this ratio, the strangeness-suppression factor in charm
photoproduction, within the LUND string fragmentation model, has been
calculated to be \( \gs =\gsvalue \), in good agreement with the $\gs$
value extracted from charm production in $e^+ e^-$ annihilation.

\section{Acknowledgements}
 
We would like to thank the DESY Directorate for their strong support
and encouragement. The remarkable achievements of the HERA machine
group were essential for the successful completion of this work and
are greatly appreciated. We would like to thank C.~Oleari for
discussions and for providing us with his latest results for \(
\epsilon (\dss ) \) and S.~Frixione and A.~Berezhnoy for providing us
with their QCD calculations.

\def\jv#1#2{#1\@\ #2}
\def\ibid#1{ibid., #1}
 
\def\CPC{\jv{Comp.\ Phys.\ Commun.}}
\def\EPJ{\jv{Eur.\ Phys.~J.}}
\def\NPPS{\jv{Nucl.\ Phys.\ Proc.\ Suppl.}}
\def\NIM{\jv{Nucl.\ Inst.\ Meth.}} 
\def\NP{\jv{Nucl.\ Phys.}}
\def\PL{\jv{Phys.\ Lett.}}
\def\PR{\jv{Phys.\ Rev.}} 
\def\PRL{\jv{Phys.\ Rev.\ Lett.}}
\def\ZP{\jv{Z.~Phys.}}

\end{document}